\documentclass[12pt,a4paper]{article}

\usepackage[utf8]{inputenc}
\usepackage[T1]{fontenc}
\usepackage{amsmath,amssymb}
\usepackage{mathrsfs,stmaryrd}
\usepackage{euscript}
\usepackage{graphicx,epic}
\usepackage[a4paper,dvipdfm,pagebackref]{hyperref}
\usepackage{color}

\definecolor{linkcol}{rgb}{0,0,0.6}
\definecolor{citecol}{rgb}{0,0.6,0}

\renewcommand*{\backref}[1]{}
\renewcommand*{\backrefalt}[4]{%
\ifcase #1 %
(Not cited.)%
\or
(Cited on page~#2.)%
\else
(Cited on pages~#2.)%
\fi}

\hypersetup
{
bookmarksopen=true,
pdftitle="A complete theory of low-energy phase diagrams for two-dimensional turbulence",
pdfauthor="Marianne Corvellec", 
pdfsubject="Article",
pdfkeywords={2D Euler equations, variational problems, equilibrium statistical mechanics,
    Lyapunov-Schmidt reduction, phase diagrams, statistical ensemble inequivalence,
    quasi-geostrophic model, Casimir functionals, tricritical point}
pdfmenubar=true,
pdfhighlight=/O,
colorlinks=true,
pdfpagemode=None,
pdfpagelayout=SinglePage,
pdffitwindow=true,
citecolor=citecol,
linkcolor=linkcol
}

\newcommand{\sgn}{\mathop{\mathrm{sgn}}}

\title{A complete theory of low-energy phase~diagrams~for~two-dimensional~turbulence
steady~states~and~equilibria}
\author{Marianne Corvellec and Freddy Bouchet}

\begin{document}

\maketitle
\begin{abstract}
For the 2D Euler equations and related models of geophysical flows, minima of energy--Casimir
variational problems are stable steady states of the equations (Arnol'd theorems). The same
variational problems also describe sets of statistical equilibria of the equations. This paper
uses Lyapunov--Schmidt reduction in order to study the bifurcation diagrams for these
variational problems, in the limit of small energy or, equivalently, of small departure from
quadratic Casimir functionals.
We show a generic occurrence of phase transitions, either continuous or discontinuous. We
derive the type of phase transitions for any domain geometry and any model analogous to
the 2D Euler equations.
The bifurcations depend crucially on $a_4$, the quartic coefficient in the Taylor expansion
of the Casimir functional around its minima. Note that $a_4$ can be related to the fourth
moment of the vorticity in the statistical mechanics framework. A tricritical point
(bifurcation from a continuous to a discontinuous phase transition) often occurs when $a_4$
changes sign.
The bifurcations depend also on possible constraints on the variational problems
(circulation, energy).   
These results show that the analytical results obtained with quadratic Casimir functionals
by several authors are non-generic (not robust to a small change in the parameters).
\end{abstract}

\section{Introduction}

Flows which are turbulent and two-dimensional are remarkable for two reasons:
First, they self-organize into large-scale coherent structures and,
second, they often display a bistable behaviour.  
Such large-scale structures (monopoles, dipoles, parallel flows)
are analogous to geophysical cyclones, anticyclones, and jets in the
oceans and atmospheres \cite{Bouchet_Sommeria:2002_JFM}.
The motion of atmospheres and oceans is almost two-dimensional indeed,
because of the following three characteristics:
i) the fluid layer has a small vertical-to-horizontal aspect ratio,
ii) the fluid layer is subject to the Coriolis force, which dominates the
viscous frictional forces, 
and iii) the water column is stably stratified over large scales,
constraining motions to be horizontal. 
In the laboratory, these characteristics can be obtained by rotating a
(shallow) cylinder filled up with water and using a forcing mechanism.
Typically, observations made on these experimental systems can shed
light on atmospheric and oceanic phenomena.
\par\smallskip\indent
The natural equations governing this type of motion are the Navier--Stokes
equations in two dimensions.  It should be noted that the large scales of
geophysical flows are highly turbulent.  Indeed,
scale analysis shows that the motion of the large scales is dominated by the
advective (also called \textit{inertial}) term; forcing and dissipation
terms are small with respect to the inertial term.
We say that the flows self-organize, precisely because the
large-scale structures are not at all determined (say, linearly) by some
external forcing.
This self-organization of the large scales is specific to 2D turbulence
\cite{Kraichnan_Phys_Fluid_1967_2Dturbulence}. Unlike in 3D turbulence,  
there is no direct energy cascade (towards small scales) but there are
an inverse energy cascade (towards largest scales) and a direct enstrophy 
cascade.  This difference stems from the conversation of different quantities
for the conservative dynamics ---then, the 2D Navier--Stokes equations
reduce to the 2D Euler equations.
If topography is included in the model,
we have instead the (inviscid) barotropic quasi-geostrophic equations.
In this inertial limit, the attractors of the dynamics are expected
to be found near a set of steady states of the inviscid equations.
\par\smallskip\indent
We have the transport of a scalar quantity $q$ by an incompressible
two-dimensional velocity; 
for the 2D Euler equations, $q$ is
the vorticity and, for the barotropic quasi-geostrophic equations, $q$ is
the potential vorticity.
We wish to predict the final state(s) of the system.
Any state verifying a functional relationship between (potential) vorticity
and streamfunction is a steady state.
Thus, there are an infinity of steady states.  How can we determine which
ones are stable?
Since $q$ is a field, the system has an infinite number of
degrees of freedom (continuous system).  A deterministic approach would
be unrealistic.  Then, we turn to statistical mechanics.  Rather than describing fine-grained 
structures (exact fields), equilibrium statistical theories of two-dimensional turbulent
flows predict ---assuming ergodicity--- the final organization of the flow at a coarse-grained level: 
a mixing entropy is maximized under the constraints that all the flow invariants be conserved
\cite{Miller:1990_PRL_Meca_Stat,Robert:1991_JSP_Meca_Stat,SommeriaRobert:1991_JFM_meca_Stat}.
There are an infinity of invariants, namely, the energy and the Casimirs;
a Casimir is any functional of the (potential) vorticity.
\par\smallskip\indent
The Miller--Robert--Sommeria theory (MRS theory, for short) predicts statistical equilibria
in terms of a functional relationship between (potential) vorticity and streamfunction.
We want to determine the large scales of (quasi) two-dimensional turbulence as
equilibria of the inviscid equations.  The analytical computation
of MRS equilibrium states would be a difficult task though: it would be about solving a
variational problem involving an infinite number of constraints.
In this paper, we present analytical and numerical computations of phase 
diagrams for a large class of equilibrium states, obtained from simpler variational
problems.
Phase transitions are very important to study, since they are associated with
major physical changes (in large-scale flow structures, as far as we are concerned)
in the system under consideration.  For instance, flows will have their structure
change as they undergo phase transitions. It is important to know whether these
are first-order (discontinuous) or second-order (continuous).
\par\medskip\indent
Indeed, simpler variational problems (taking into account only a few constraints) were shown to
give access to some classes of MRS equilibria \cite{Bouchet:2008_Physica_D}.  For instance,
one such class is the one for which $q = f(\psi)$ is linear (or affine).
An example of using statistical mechanics for predicting and describing
real turbulent flows can be found in \cite{Bouchet_Simonnet_2009PRL} and references therein.
Bifurcations between stable steady solutions of 2D Euler
are found to occur when varying the domain shape, the nonlinearity of $f(\psi)$, or the energy.
This suggests that a general theory of phase transitions for 2D
and geophysical flows should be looked for ---it is not available at the present day.
Only instances of such phase transitions have been reported in the literature.
Note that key results regarding statistical ensemble inequivalence, encompassing the case of a
nonlinear equation $q = f(\psi)$, were presented in~\cite{Ellis_Haven_Turkington_2002_Nonli}.
In this paper, we present new analytical results on phase transitions related to
the nonlinearity of $f(\psi)$.
We obtain a complete theory of phase diagrams for two-dimensional turbulence equilibria
and steady states in the low-energy limit.
\par\medskip\indent
The simpler variational problem we consider writes
\begin{equation}\label{micro-varpb}
C_s(E, \Gamma) = \min_{q} \left\{ \int_{\mathcal D} s(q) \ | \ \mathcal{E}[q] = E\,, {\mathit \Gamma}[q] = \Gamma \right\}.
\end{equation}
The function $s(q)$ is assumed strictly convex.  In thermodynamics, the
microcanonical problem is a two-constraint variational problem where the thermodynamical
potential to be maximized is called the entropy.  We can draw an analogy with
\eqref{micro-varpb}, where our Casimir functional $\int_{\mathcal D} s(q)$ acts
as the opposite of an entropy. 
We give the expressions\footnote{See \eqref{eq:btQG} and section \ref{sec:def-res}
for a definition of the fields $\psi$ and $h$.} of
\begin{align}
\mathcal{E}[q] & = -\frac{1}{2} \int_{\mathcal D} \psi(q-h)\,, \;
\text{the kinetic energy, and} \notag \\
{\mathit \Gamma}[q] & = \int_{\mathcal D} q\,, \; \text{the circulation.} \quad\notag
\end{align}
So we call \eqref{micro-varpb} microcanonical, 
in analogy with usual thermodynamics. 
Note that this variational problem corresponds to (CVP) in~\cite{Bouchet:2008_Physica_D}
(see this reference about the relationship between the solutions to our
variational problem and the actual MRS statistical equilibria).
For given values of the constraints $E$ and $\Gamma$, the $q$ fields solving \eqref{micro-varpb}
are microcanonically stable equilibria.  This is a sufficient condition for
their dynamical stability \cite{Arnold_1969}.  Indeed, let us
consider a functional which is conserved by the dynamics.  This functional can be a linear
combination of a Casimir and of the energy (`energy--Casimir functional').
The point is the following: if the system lies at a nondegenerate extremum
of this invariant, then it cannot go away from this point.
\par\medskip\indent
The paper is organized as follows: In section \ref{sec:def-res}, we define the
quantities and notions in use throughout the work and we give the general results. 
It appears that phase transitions can be characterized through the bifurcation
analysis of scalar equations, the latter acting as normal forms.
The technical derivation of the results is given in the various appendices.
In section \ref{sec:example}, we apply the general results to a particular case,
in a rectangular domain. Equilibria are computed numerically using the pseudo-arclength
continuation method. Finally, in section \ref{sec:applns}, we suggest some physical
applications, offering a clear motivation for this theoretical work.

\section{Definitions and general results\label{sec:def-res}}

The system we consider is that of the barotropic quasi-geostrophic equations, which model
the 2D dynamics of one oceanic or atmospheric layer:
\begin{equation}\label{eq:btQG}
\partial_t q + \mathbf{u} \cdot\! \nabla q = 0 \;;\quad \mathbf{u} = {\bf e}_z \times \nabla \psi
\;;\quad q = \Delta \psi + h
\end{equation}
where ${\mathbf u}$ denotes the (two-dimensional) velocity field, $\psi$ the
streamfunction (defined up to a constant),
$q$ the potential vorticity (in vorticity units), and $h$ an equivalent topography.
The boundary condition is $\psi = 0$ on $\partial \mathcal D$, 
where $\mathcal D$ is a simply connected domain in two dimensions.
The natural scalar product for the fields at play is denoted by
$\langle q_1 q_2 \rangle := \int_{\mathcal D} q_1 q_2$.
\par\smallskip\indent
The inviscid dynamics \eqref{eq:btQG} corresponds to a limit of infinite
Reynolds number.
Although the number of degrees of freedom is infinite in a
turbulent flow, the formation of large-scale structures indicates that
just a few effective degrees of freedom should be enough to characterize
the flow.  
In this paper, we describe a class of steady states of \eqref{eq:btQG},
and the phase transitions which they undergo, through scalar bifurcation equations.
The stability of these steady states can be established statistically
(thermodynamically), implying dynamical stability (Arnol'd's theorems).
In the equilibrium statistical-mechanical context, we deal with
phase transitions.  In the dynamical system context, we deal with bifurcations.
Here, we use technical tools of applied bifurcation theory, namely, Lyapunov--Schmidt reduction,
to characterize the phase transitions.
Hence, we can determine the continuous or discontinuous nature of phase transitions in a
general framework.

\subsection{Relaxed variational problems}

To compute statistical equilibria, which are, again, particular steady states
of the dynamics \eqref{eq:btQG}, we solve the \textit{microcanonical variational problem}
\eqref{micro-varpb}, as announced in the introduction.
In this paper, we restrict our attention to even functions $s(q)$.  
Indeed, there are many situations where the $q \mapsto -q$ symmetry applies.
If $q$ is a solution to \eqref{eq:btQG}, then $-q$ is also a solution to
\eqref{eq:btQG}.  In real flows, the $q \mapsto -q$ symmetry could be broken
by a nonsymmetric forcing or by a nonsymmetric initial distribution of
(potential) vorticity.
Say that $s$ can be written as the expansion
\begin{equation}\label{eq:chap2-normal}
s(q) = \frac{1}{2} q^2 - \sum_{n \geq 2} \frac{a_{2 n}}{2 n} q^{2 n}.
\end{equation}
Assuming that the Lagrange multiplier rule applies ($q$ regular enough),
there exists a couple $(\beta, \gamma) \in \mathbb{R}^2$ such that
solutions of \eqref{micro-varpb} are stationary points of
\begin{equation}\label{G_functional}
\mathcal{G}[q] = \int_{\mathcal D} s(q) + \beta \mathcal{E}[q] + \gamma {\mathit \Gamma}[q].
\end{equation}
We call this functional the Gibbs free energy, in analogy with usual thermodynamics.
The variational problem dual to \eqref{micro-varpb}, i.e.,
\begin{equation}\label{relaxed-varpb}
G(\beta, \gamma) = \min_{q} \left\{ \mathcal{G}[q] = \int_{\mathcal D} s(q) + \beta \mathcal{E}[q] + \gamma {\mathit \Gamma}[q] \right\},
\end{equation}
is referred to as the \textit{grand canonical variational problem}.  Because it is relaxed
(unconstrained), it is more easily tractable.
\par\medskip\indent
We shall consider another energy--Casimir variational problem, namely,
the \textit{canonical variational problem}:
\begin{equation}\label{canon-varpb}
F(\beta, \Gamma) = \min_{q} \left\{\mathcal{F}[q] = \int_{\mathcal D} s(q) + \beta \mathcal{E}[q] \ |
\ {\mathit \Gamma}[q] = \Gamma \right\}.
\end{equation}
It is the problem of minimizing the Helmholtz free energy with fixed circulation $\Gamma$.

\subsection{Ensemble inequivalence\label{sub:ens-ineq}}

For given values of the constraints $E$ and $\Gamma$, the $q$ fields solving \eqref{micro-varpb} are
statistical equilibria.
As introduced in the previous subsection,
$\beta$ and $\gamma$ are the Lagrange multipliers associated with the energy and circulation constraints, respectively.
For all couples $(\beta, \gamma)$, minima $G(\beta, \gamma)$ are also minima $C_s(E(\beta, \gamma), \Gamma(\beta, \gamma))$.
But some minima $C_s(E, \Gamma)$ may correspond to stationary points of \eqref{G_functional} which are
not minima of \eqref{G_functional}.  These are classical results (see any textbook on convex optimization).
When $E(\beta, \gamma)$ and $\Gamma(\beta, \gamma)$ do not span their entire accessible range ($E \in \mathbb{R}_{+}$,
$\Gamma \in \mathbb{R}$) as $(\beta, \gamma)$ is varied,
the microcanonical ensemble and the (dual) grand canonical ensemble are said to be inequivalent.
Then, some microcanonical solutions are not obtained as grand canonical solutions.
\par\smallskip\indent
This feature is typical of long-range interacting systems.  We use the
term `long-range interactions' as in, for instance, \cite{Bouchet_Barre:2005_JSP}
and \cite{Campa_et_al_PhysRep2009}: for a
system in space dimension D, the interaction potential between particles
separated by a distance $r$ goes like $r^{-\alpha}$, as $r \to \infty$,
with $\alpha \leq$ D.  The interaction is `non-integrable'.
From the expression of kinetic energy for 2D Euler, the coupling between vorticity
at point $\mathbf{r}$ and vorticity at point $\mathbf{r}'$ appears to be logarithmic,
hence not integrable. Thus, the vorticity at a given point is coupled with the vorticity
of any other point of the domain, not only of neighbouring points.
In addition to 2D turbulence, long-range interacting systems include
self-gravitating systems in astrophysics and some models in plasma physics.
In short-range interacting systems, the different
statistical ensembles are used interchangeably, because they are usually equivalent.
\par
\begin{figure}[htp]
\begin{center}
\includegraphics[width=5.5in]{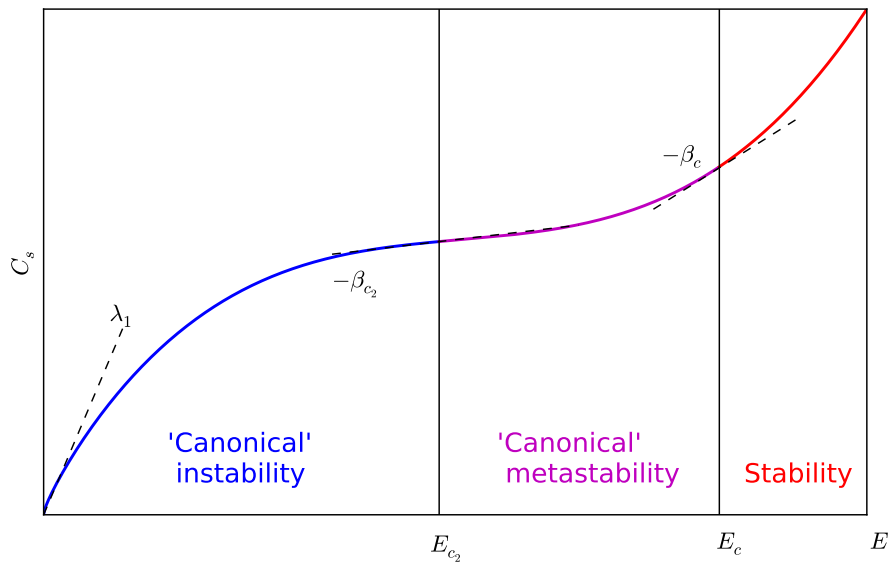}\\
\caption{\small $C_s(E)$ showing a range of `canonical stability' for $E > E_c$: minima $C_s(E)$ are
minima $F(\beta)$; a range of `canonical metastability' for $E \in [E_{c_2}, E_c]$: minima $C_s(E)$
can be obtained as local minima of the canonical functional; a range of `canonical instability'
for $E \in \: ]0, E_{c_2}[$:
minima $C_s(E)$ can be obtained as local maxima of the canonical functional.}
\label{fig:CsofE}
\end{center}
\end{figure}

\medskip\indent
Let us illustrate the idea of ensemble inequivalence with a schematic picture.
For the sake of simplicity, let us discard the circulation constraint.
The microcanonical solutions are described by $C_s(E)$.  If the caloric curve
$\beta(E) = -C_s'(E)$ is monotonically decreasing, i.e., if $C_s(E)$ is convex,
the microcanonical solutions can all be obtained as canonical solutions:
the two ensembles are equivalent.  If the caloric curve is increasing over a
certain range (negative specific heat, in thermodynamics terms), there is a
range of ensemble inequivalence.
The canonical ensemble is equivalent to the microcanonical one only
over the range for which $C_s(E)$ coincides with its convex envelope (range
$E > E_c$ on Figure \ref{fig:CsofE}): solutions $C_s(E)$ are solutions $F(\beta)$.
$C_s(E)$ has an inflexion point
at $E = E_{c_2}$ (we shall use this notation in subsection \ref{sub:gamzer}).
It is a canonical spinodal point.
We refer the reader to \cite{Bouchet_Barre:2005_JSP} for a systematic classification
of all these singularities.

\subsection{Poincar\'e inequality\label{sub:Poin-ineq}}

The Poincar\'e inequality comes in handy to establish a sufficient condition for convexity;
it is natural to begin with the study of the convexity of $\mathcal{G}[q]$ \eqref{G_functional}.
Indeed, it is readily noted that if $\mathcal{G}[q]$ is strictly convex,
it has a unique stationary point, which is then the (unique) solution of \eqref{micro-varpb}.
Since ${\mathit \Gamma}[q]$ is a linear form, it is sufficient to investigate the convexity of
the Helmholtz free energy functional $\mathcal{F}[q] = \int_{\mathcal D} s(q) + \beta \mathcal{E}[q]$.
Since $\mathcal{E}[q]$ is convex, $\mathcal{F}[q]$ is strictly convex if $\beta \geq 0$.
\par\smallskip\indent
If $\beta < 0$, we need to study the sign of the second-order variation of $\mathcal{F}$,
denoted by $\delta^2 \mathcal{F}$, and defined
through $\mathcal{F}[q + \delta q] - \mathcal{F}[q] = \delta \mathcal{F}[q] + \frac{1}{2} \delta^2 \mathcal{F}[q] + o(\delta q^2)$.
We get
\begin{align}
\delta^2 \mathcal{F}[q] = \int_{\mathcal D} s''(q) \delta q^2 - \beta \int_{\mathcal D} \delta\psi \delta q.
\end{align}
We make use of the Poincar\'e inequality:
\begin{align}\nonumber
\int_{\mathcal D} \delta\psi \delta q \geq -\frac{1}{\lambda_1}  \int_{\mathcal D} \delta q^2,
\end{align}
where $-\lambda_1 < 0$ is the greatest (smallest, in absolute value) eigenvalue of the Laplacian on $\mathcal{D}$.
Indeed (let us recall the classical proof), introducing
the orthonormal Laplacian eigenbasis $\{e_i({\bf r})\}_{i \geq 1}$, i.e.,
\begin{equation}
\Delta e_i({\bf r}) = - \lambda_i e_i({\bf r}) ,\quad \int_{\mathcal D} e_i e_j = \delta_{ij} , \quad 0 < \lambda_1 < \lambda_2 < \ldots,
\end{equation}
with $e_i = 0$ on $\partial \mathcal D$ for all $i \geq 1$.
All fields may be decomposed in this basis:
\begin{align}
\delta\psi({\bf r}) &= \sum_i \delta\psi_i e_i({\bf r}), \notag \\
\delta q ({\bf r}) &= \delta (\Delta \psi ({\bf r}) + h({\bf r})) = \delta \Delta \psi ({\bf r}) = -\sum_i \lambda_i \delta \psi_i e_i({\bf r})  \notag \\
& = \sum_i \delta q_i e_i({\bf r}).
\end{align}
Therefore,
\begin{align}
\int_{\mathcal D} \delta\psi \delta q &= - \sum_i \frac{\delta q_i^2}{\lambda_i} \geq - \frac{1}{\lambda_1}  \sum_i \delta q_i^2 =
- \frac{1}{\lambda_1} \int_{\mathcal D} \delta q^2. \notag
\end{align}
So
\begin{align}\label{secondvar}
\delta^2 \mathcal{F}[q] \geq \int_{\mathcal D} \left(s''(q) + \frac{\beta}{\lambda_1} \right) \delta q^2 \geq
\left(s''_m+ \frac{\beta}{\lambda_1} \right) \int_{\mathcal D} \delta q^2,
\end{align}
for $\beta < 0$, where $s''_m := \min_{\mathbf{r} \in \mathcal{D}} \{\min_q s''(q(\mathbf{r}))\}$.
If $\beta > - s''_m \lambda_1$, $\mathcal F$ is strictly convex, and so is $\mathcal G$.
There is a unique solution to \eqref{relaxed-varpb} and, hence, a unique solution to \eqref{micro-varpb}.
\par\medskip\indent
The conditions $\beta \geq 0$ and $- s''_m \lambda_1 < \beta < 0$ are the hypotheses for
the first and second Arnol'd theorems, respectively, on Lyapunov stability.  In both cases,
the sufficient condition is that $\delta^2 \mathcal{F}$ be positive-definite
\cite{Michel_Robert_1994_JSP_GRS}.
We can conclude that for $\beta > - s''_m \lambda_1$, the grand canonical ensemble is
equivalent to the microcanonical ensemble.
In the grand canonical ensemble, phase transitions may occur only for
$\beta \leq - s''_m \lambda_1$, where solutions to \eqref{relaxed-varpb}
may cease to be unique or cease to exist.
\par\medskip\indent
The stationary points of $\mathcal{G}$ \eqref{G_functional} are the $q$ fields for which the first-order
variation of $\mathcal{G}$ vanish, i.e.,
\begin{equation}\label{qfpsi}
s'(q) - \beta \psi + \gamma = 0. 
\end{equation}
Since $s(q)$ is strictly convex, $s'(q)$ is strictly increasing, so its inverse $(s')^{-1}(q)$ is
well-defined (and strictly increasing).  We have
\begin{equation}\nonumber
q = (s')^{-1}(\beta \psi - \gamma).
\end{equation}
From \eqref{eq:chap2-normal},
the Taylor expansion of $(s')^{-1}$ around $0$ reads $(s')^{-1}(x) = x + a_4 x^3 + o(x^4)$.
Then, the term in $a_4$ is the lowest-order nonlinear contribution to $(s')^{-1}(x)$.

\subsection{Phase diagram for $\gamma = 0$\label{sub:gamzer}}

It is more straightforward to study a symmetric problem first and, afterwards,
to study the effect of breaking the symmetry.  Therefore, we begin with the
case $\gamma = 0$ so that \eqref{relaxed-varpb} is symmetric with respect to
$q \mapsto -q$.  The corresponding constrained variational problem is the
\textit{grand microcanonical variational problem with $\gamma = 0$}, i.e.,
the minimization of $\int_{\mathcal D} s(q)$ with fixed energy.
We find that the grand canonical ensemble with $\gamma = 0$
is equivalent to the grand microcanonical (only energy-constrained) ensemble
if $a_4 \leq 0$.  It is not the case if $a_4 > 0$.
\par\smallskip\indent
We have denoted the first (largest-scale) Laplacian eigenmode by $e_1$.
As long as the topography field $h$ is orthogonal to $e_1$,
we find the following results, for the grand canonical ensemble with $\gamma = 0$:
\begin{itemize}

\item for $a_4 \leq 0$, there is a second-order phase transition at $\beta = -\lambda_1$: the solution goes continuously
from a trivial state (zero energy, uniform vorticity) to a state dominated by $e_1$; 
\item for $a_4 > 0$, $a_4$ small enough, there is a first-order phase transition at
$\beta = \beta_c(a_4) \in\ ]-\lambda_1, -\lambda_1 s''_m[$: the solution goes discontinuously from a trivial state ($E = 0$) to a state dominated by $e_1$ ($E = E_c(a_4) > 0$).  The energy range accessible by grand canonical solutions (with $\gamma = 0$) displays a gap
$]0, E_c(a_4)[$.

\end{itemize}
Systems with symmetry display a richer phenomenology of phase transitions,
especially regarding second-order phase transitions \cite{Bouchet_Barre:2005_JSP}.
So it is not surprising to find a second-order phase transition line here.
\par\smallskip\indent
In the grand microcanonical ensemble with $\gamma = 0$, we find that
\begin{itemize}
\item there is no phase transition at low energy (we cannot tell what happens at high energy);
\item at nonzero low energy, the solution is a state dominated by $e_1$;
\item for $a_4 > 0$, states of lowest energy ($E \in [0, E_{c_2}(a_4)]$) have negative specific heat.

\end{itemize}
\par\medskip\indent
What is the method for deriving these results?  We explain it qualitatively here
and give the technical details in appendix \ref{app:gdcan-sol}.
When solving \eqref{micro-varpb}, the quadratic part of $s$ comes into play at lowest (linear) order
in $E$~\cite{Bouchet_Simonnet_2009PRL}.  Therefore, in the low-energy limit, it is the dominant contribution.
Also, at lowest order, the solution is along $e_1$, the largest-scale eigenmode.
The next order brings into
play the small parameter $a_4$, referred to as the nonlinearity, for short.
We may always write $q = A e_1 + q'$ with $A \in \mathbb{R}$ and $q'$ orthogonal to $e_1$.
We see $q'$ as a perturbation to the lowest-order solution $\pm A e_1$
and assume it admits an asymptotic
expansion in (powers of) $A$.  This will lead to an asymptotic expansion in $A$ for the Gibbs free energy,
i.e., a normal form describing the phase transitions in a neighbourhood of $a_4 = 0$. 
The idea is to minimize ${\mathcal G}$ with respect to $q'$ first, then with respect to $A$.
We expect the symmetries at play to show in this normal form.
\par\smallskip\indent
Thus, we have reduced the infinite-dimensional variational problem ---or equation for the
stationary points \eqref{qfpsi}--- to a scalar equation, the
\textit{bifurcation equation} \cite{chow1982methods}.
This is called Lyapunov--Schmidt reduction.
We have solved the unconstrained (or grand canonical) variational problem
in a symmetric case.  We find a tricritical point.

\subsection{Tricritical point}

\begin{figure}[htp]
\begin{center}
\includegraphics[width=4.in]{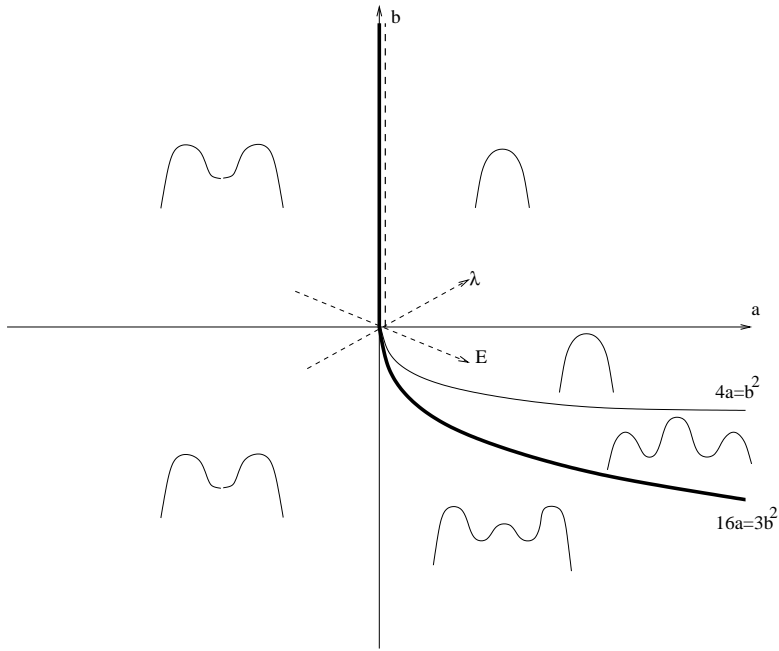}\\
\caption{\small The `canonical' tricritical point is at $(a,b) = (0,0)$.  The curve $(4a = b^2, b < 0)$ corresponds
to the appearance of three local maxima.  The bold curve $(16 a = 3 b^2, b < 0)$ is a first-order phase transition line.  The bold-dashed
curve is a second-order phase transition line.  Here, `canonical' simply refers to a relaxed ensemble with respect
to a constrained one.  Figure from \cite{Bouchet_Barre:2005_JSP}.} \label{fig:tricritical}
\end{center}
\end{figure}

A tricritical point is a point where a second-order phase transition meets a first-order one.
We have predicted the phase diagrams in the vicinity of $(\beta, a_4) = (-\lambda_1, 0)$.
For $\gamma = 0$ and $h_1 = 0$, the normal form \eqref{GofA-g} is 
\begin{equation}\nonumber
G_0(A) = \frac{1}{2} \left(1 + \frac{\beta}{\lambda_1}\right) A^2 - \frac{a_4}{4} \langle e_1^4 \rangle A^4 + o(A^5).
\end{equation}
We can readily relate this expression to the normal form $s_{a,b}(m) = -m^6 - 3 b m^4/2 - 3 a m^2$.
This normal form is used in the context of constrained variational problems in
\cite{Bouchet_Barre:2005_JSP}.   Note that $s_{a,b}(m)$ is to be maximized, and hence, solutions are maximizers there.
Then, the identification of coefficients is to be done between $s_{a,b}$ and $-G_0$.
The typical behavior of $s_{a,b}$ and the associated transition lines are shown on Fig. 6 of this reference, reproduced
below (our figure \ref{fig:tricritical}).
\par\smallskip\indent
If $a > 0$ and $b > 0$, then $s_{a,b}$ is concave and there is only one maximizer, namely $m =0$.  We can see that $m = 0$ is always
a critical point.  The other possible stationary points are such that $m^4 + b m^2 + a = 0\,.$
For $b \geq 0\,,$ a pair of maxima appears as $a$ becomes negative, hence the second-order phase transition at $(a = 0, b \geq 0)$.
A pair of minima and a pair of (local) maxima appear as $|b| \geq 2 \sqrt{a}$ in the quarter plane $(a \geq 0, b \leq 0)$.
There is a first-order phase transition when these maxima reach $s_{a,b}(m = 0) = 0$ as $a$ and $b$ decrease.
It is found to occur for $16a = 3b^2$.
\par\smallskip\indent
Parameters $a$ and $b$ are identified with $(1 + \frac{\beta}{\lambda_1})/6$ and $-a_4 \langle e_1^4 \rangle/6$
respectively.  Therefore, $(\beta, a_4) = (-\lambda_1, 0)$ is a tricritical point in the grand canonical ensemble with
$\gamma = 0$.  The normal form sketched in the various areas of the phase diagram (Figure \ref{fig:tricritical})
should be identified with the opposite of $G_0(A)$.  Stationary points near $A = 0$ are found to be $A_0 = 0$ for all
$(\beta, a_4)$; in addition,
\begin{align}\label{eq:Apm-g}
A_{\pm} &= \pm \sqrt{\frac{\beta + \lambda_1}{\lambda_1 a_4 \langle e_1^4 \rangle}} + O\left(\left(1 + \frac{\beta}{\lambda_1}\right)^{3/2}\right)
\end{align}
are stationary points when $\beta + \lambda_1$ and $a_4$ have the same sign.
\begin{itemize}

\item[--] For $\beta > -\lambda_1$, $A_0$ is a local minimum. 
\item[--] For $\beta < -\lambda_1$ and $a_4 < 0$, $\{A_{\pm}\}$ are local minima originating from symmetry-breaking:
there is a second-order phase transition at $(\beta = -\lambda_1, a_4 < 0)$.
\item[--] For $\beta > -\lambda_1$ and $a_4 > 0$, $\{A_{\pm}\}$ are local maxima.  Then, minima far away from $A = 0$ have
to exist, for $G_0(A)$ has a lower bound, owing to the convexity of $s(q)$.  These, say, `nonlocal' minima cannot be obtained
perturbatively.
\item[--] For $\beta < -\lambda_1$ and $a_4 > 0$, $A_0$ is a local maxima; it is the only
stationary point obtained perturbatively.
Solutions have to be the above-mentioned nonlocal minima.  So there has to be a first-order phase
transition at $\beta > -\lambda_1$, where the solution jumps from $A_0$ to the `nonlocal' minima.

\end{itemize}
Since we also know (see subsection \ref{sub:Poin-ineq}) that $A_0$ is the only solution for $\beta > -\lambda_1 s''_m$,
the first-order phase transition is a line $\beta_c(a_4 > 0)$ such that $\beta_c(a_4) \in ]-\lambda_1, -\lambda_1 s''_m(a_4)[$.

\subsection{Phase diagram for constant circulation}

Now we solve the canonical variational problem: the energy constraint is relaxed,
the circulation is fixed at a low value.
We find interesting phase transitions, where the flow structure completely changes.  For elongated rectangular
domains (aspect ratio $\tau > \tau_c$), we recover the showing up of a dipolar structure
(contribution from mode $e_1'$), while for square-like domains
($\tau < \tau_c$), we recover that of a central monopole with counter-circulating cells at the
corners (contribution from mode $e_*$) \cite{Chavanis_Sommeria_1996JFM,Venaille_Bouchet_PRL_2009}.  The novelty here is to distinguish
between a first-order transition and a second-order one, depending on the sign of the nonlinearity in $q=f(\psi)$, at
zero circulation. 
First of all, we restrict our study to the case of zero circulation ($\Gamma = 0$), bringing
symmetry to our system.  As noted earlier, systems with symmetries are well known to display
a richer phenomenology of phase transitions.  We obtain phase diagrams with tricritical points, again.
Thus, results in the microcanonical ensemble (with $\Gamma = 0$) can be deduced the same way as in subsection
\ref{sub:gamzer}: for  $a_4 \leq 0$, no singularity of $C_s(E)$; for $a_4 > 0$, canonical spinodal point, negative specific heat
for the lowest-energy states.
\par\smallskip\indent
In the linear case ($a_4 = 0$), the constrained canonical problem was transformed into a
tractable equivalent unconstrained problem~\cite{Venaille_Bouchet_PRL_2009}.  We use the same trick,
as detailed in appendix \ref{app:canon-sol}.
In appendix \ref{app:LSred}, we detail the computation of the $\Gamma = 0$ solutions.
Because we linearize \eqref{qfpsi}, it is natural to recover the same critical values
(collectively denoted by $\lambda_c$)
and neutral directions (collectively denoted by $e_c$) as in the linear case
($a_4 = 0$), which was investigated by \cite{Chavanis_Sommeria_1996JFM,Venaille_Bouchet_PRL_2009}.
\par\smallskip\indent
At small but nonzero circulation, we lose the second-order phase transition to symmetry-breaking, but
then we have metastable states (of which stability can be made as close as wanted to that of the equilibrium,
as the circulation tends to zero).  In the square-like case, we can be in the presence of three qualitatively different states
(stable or metastable).
\par\smallskip\indent
Let us consider the phase space $(\beta, a_4)$.  Right to the first-order phase transition line, the solution is a weak monopole
(the amplitude $A$ of $e_c$ is very close to $0$).  As the first-order phase transition line is crossed, $|A|$ jumps to a larger
value, giving a different structure to the solution flow.  For example, in case {\bf ii)}, the transition to a dipolar contribution
is abrupt in the upper half-plane, while it is smooth in the lower half-plane, with a canonical metastable state showing up (local minimum).
The reader is referred to Figure \ref{fig:ph-diag-canii}.
\par
\begin{figure}[htp]
\begin{centering}
\input{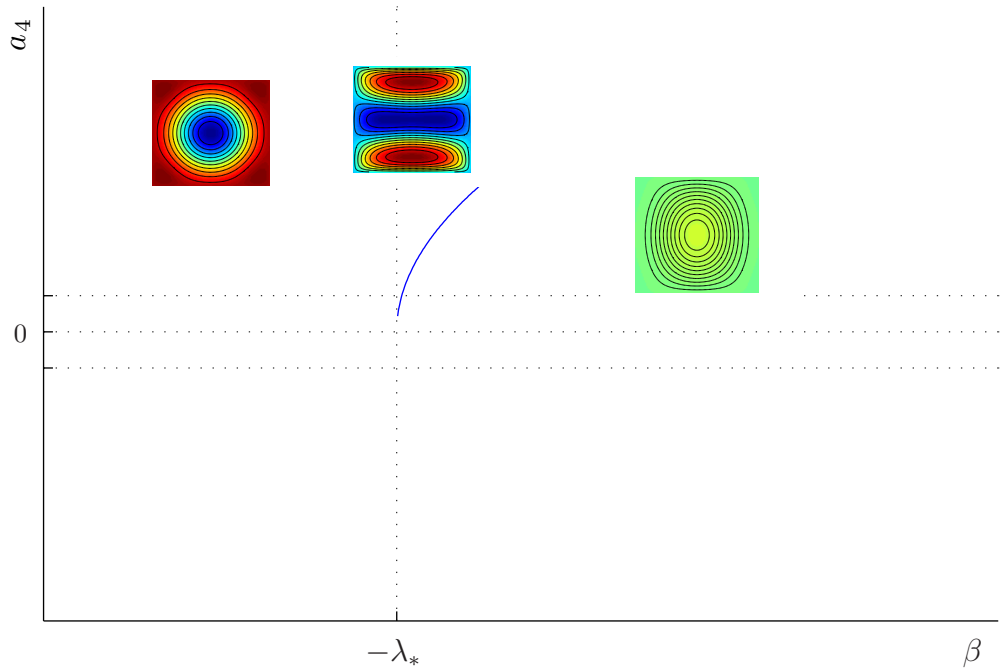}
\caption{Schematic phase diagram in the canonical ensemble for a rectangle of aspect ratio $1.1$
(case {\bf i)}).  The blue curve represents the first-order phase transition line.  This first-order phase transition line ends at a (microcanonical) critical point, which is located above and close to the point $(-\lambda_*, 0)$.  Insets show vorticity fields.
The color scale shows negative (resp. positive) values in blue (resp. red); the black contours are ten iso-vorticity lines on each plot.  From left to right: equilibrium dominated by $e_*$; equilibrium consisting of equivalent contributions from $e_*$
and the low-circulation monopole; low-circulation monopole.} \label{fig:ph-diag-cani}
\end{centering}
\end{figure}

\smallskip\indent
Figure \ref{fig:ph-diag-cani} shows a schematic phase diagram for case {\bf i)}.  Equilibrium states of the left-hand side have
different topologies, depending on the relative contributions of the monopole and $e_*$.  The contribution of the monopole
is determined by $|\Gamma|$, that of $e_*$ by $|A|$.  For certain values of $\Gamma$, there is a region in the
left-hand-side neighborhood of $(\beta, a_4) = (-\lambda_*, 0)$ where the two contributions have the same order of
magnitude, yielding a tripolar structure for the equilibrium states.  At large $|A|$ (i.e., very negative $\beta$, at given $a_4$),
only $e_*$ contributes to the structure of the equilibrium states.

\par\smallskip\par

\par\medskip\indent
We wish to emphasize that the canonical ensemble may be relevant to geophysical applications, since
the two regimes of known bistable systems have different energies.  The area of phase diagram near the discontinuous
transition should be that of interest, when investigating stochastically induced transitions.

\newpage
\section{Example: Rectangular domain\label{sec:example}}

In this section, we apply our general results to the (simple) case of
a rectangular domain of area unity: $\mathcal D = \{ (x,y) \in [0, \tau^{1/2}] \times [0, \tau^{-1/2}] \}$
with $\tau \geq 1$.
We choose a function $s(q)$ such that
\begin{align}\notag
s'(q) &= \left(\frac{1}{3} - 2 a_4\right) \tanh^{-1} (q) +  \left(\frac{2}{3} + 2 a_4\right) \sinh^{-1} (q)
\end{align}
with $a_4 \in [-1/3\,, 1/6]$ so that $s(q)$ is convex, as required.
Bound $a_4 = -1/3$ corresponds to $q = \tanh(\beta \psi)$ (two-level vorticity distribution
$\{\pm 1\}$ in the MRS theory), while bound $a_4 = 1/6$ corresponds to
$q = \sinh(\beta \psi)$ (three-level vorticity distribution $\{\pm 1, 0\}$ in the MRS theory).
We have $a_6 = a_4/4 - 7/60$.  We take $h = 0$.
\par\smallskip\indent
The corresponding steady states are computated numerically by a method of continuation,
namely, pseudo-arclength continuation.
Pseudo-arclength continuation is well-suited for computing solution branches which undergo bifurcations.
Our continuation parameters are the control parameters, $\beta$ and $a_4$.

\subsection{Solutions for $\gamma = 0$}

\begin{figure}[hbp]
\includegraphics[width=\textwidth]{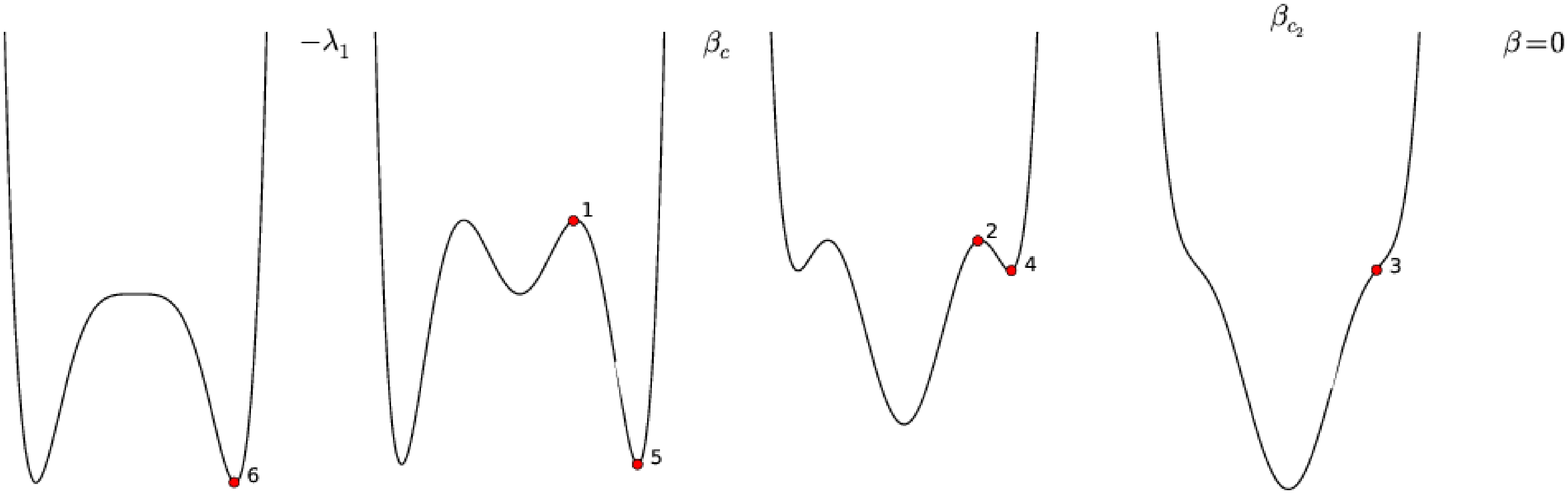}\\
\caption{\small Schematic representation of $G_0(A)$ for a given value of $a_4 > 0$
and different values of $\beta$, in increasing order from $\beta < -\lambda_1$ (left)
to $\beta_{c_2}$ (right).  The red bullets (stationary points) are numbered according
to the path taken by the continuation computation: `1' and `2' are $A_{+}$; at `3'
we bifurcate into the `nonlocal' minimum of $G_0(A)$.}
\label{fig:cont-path}
\end{figure}

We begin with $\gamma = 0$ and $\tau = 1$ (square domain).
We solve $\Delta \psi = (s')^{-1}(\beta \psi)$ \eqref{qfpsi} in $\psi$, that is,
we compute the stationary points of $\mathcal{G}_0$.
Thanks to the parity symmetry, we may restrict our study to the domain $A \geq 0$.
For a given $a_4 > 0$, $A_{+}$ \eqref{eq:Apm-g} is the local maximum.  If we increase $\beta$ from $-\lambda_1^{+}$
up to $\beta_{c_2}$, we can bifurcate into the `nonlocal' minimum of $G_0(A)$,
as represented on Figure \ref{fig:cont-path}.
Thinking of $A$ as an order parameter, there is a fold bifurcation at $\beta = \beta_{c_2}(a_4)$.

\par
\begin{figure}[htp]
\begin{center}
\includegraphics[width=4.0in]{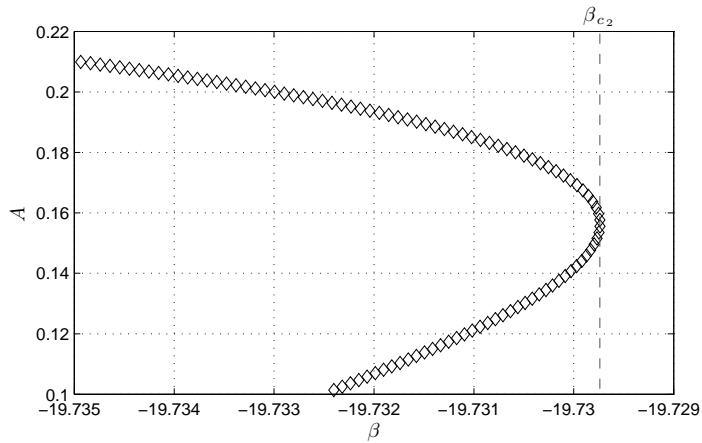}\\
\caption{\small $A(\beta)$ for $a_4 = 0.015$.} \label{fig:fold}
\end{center}
\end{figure}

\medskip\indent
In a square domain ${\mathcal D}$, $\lambda_1 = 2 \pi^2 \approx 19.7392$ and
$e_1(x,y) = 2\sin(\pi x) \sin(\pi y)$.
We start at $(\beta, a_4) = (-\lambda_1 + 0.006, 0.015)$ with solution guess
\begin{equation}\nonumber
\psi = -\frac{A_{+}}{\lambda_1} e_1.
\end{equation}
Let us denote by $A_{\mathrm{comp}}$ the scalar product of the (computed) solution $q$ with mode
$e_1$.  $|A_{\mathrm{comp}} - A_{+}|$ must scale like $A_{+}^3$. 
We check that we caught the proper solution branch by verifying this scaling relation.
Figure \ref{fig:fold} shows $A_{\mathrm{comp}}$ as a function of $\beta$, displaying
the expected fold bifurcation.

\begin{figure}[htp]
\begin{center}
\includegraphics[width=4.0in]{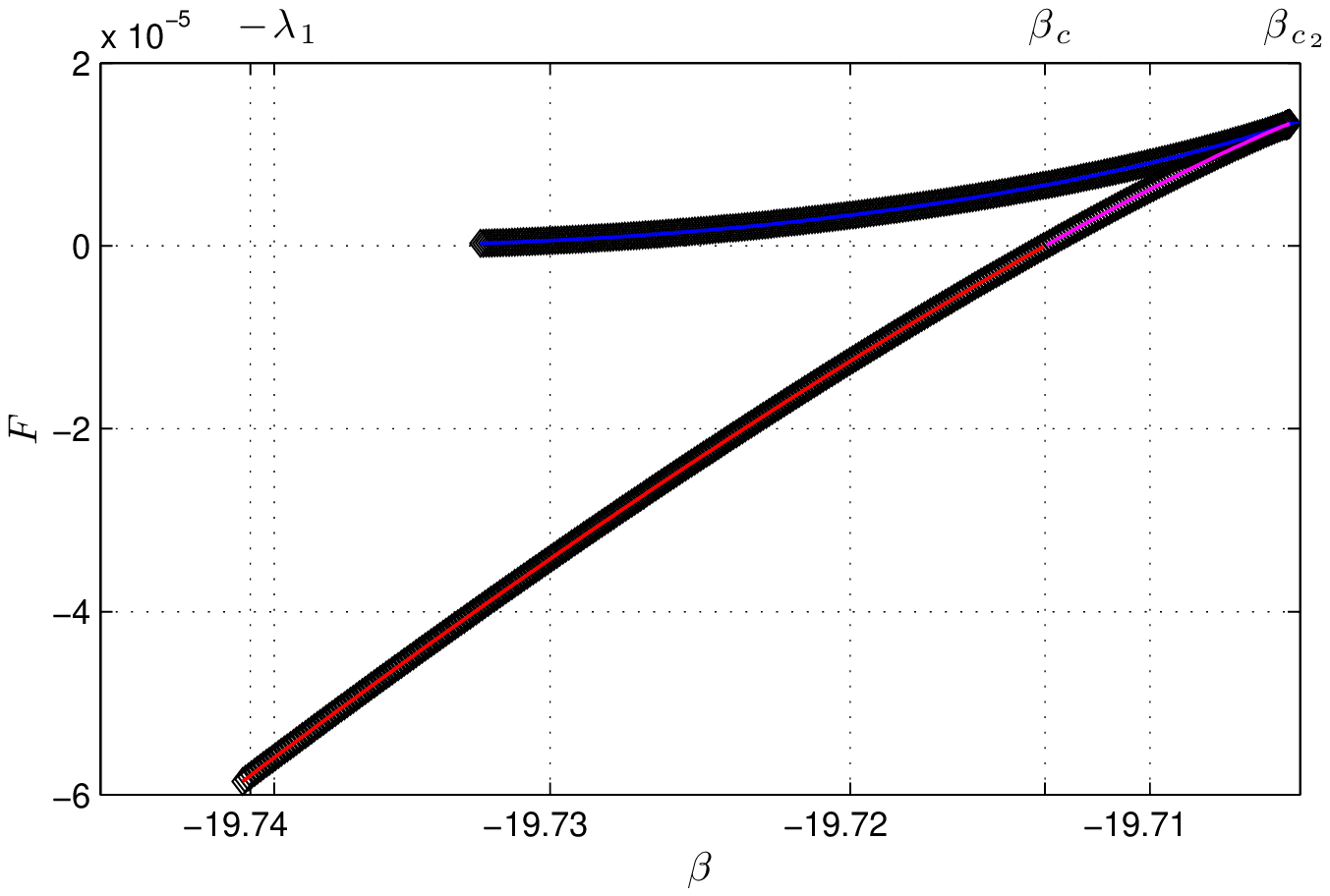}\\
\caption{\small $G_0(\beta)$ for $a_4 = 0.030$.  Local maxima are in blue; local minima in magenta;
global minima in red.} \label{fig:Fofbeta}
\end{center}
\end{figure}

\begin{figure}[htp]
\begin{centering}
\includegraphics[width=4.0in]{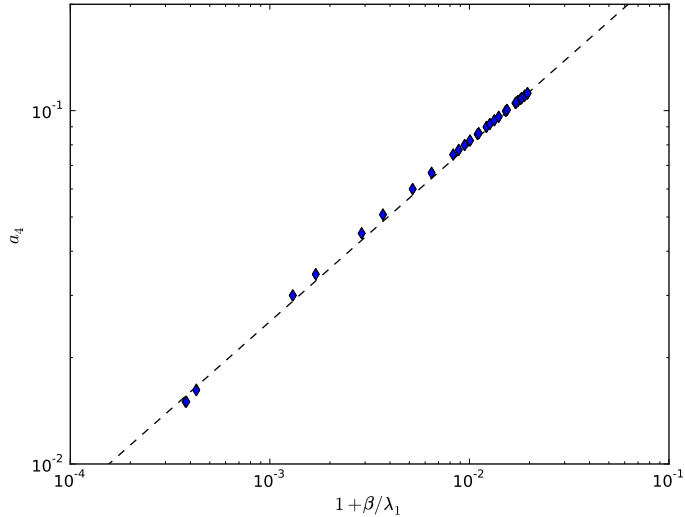}\\
\caption{\small $\log(a_4)$ as a function of $\log(1 + \beta_c(a_4)/\lambda_1)$.
The cyan and blue points, almost superimposing, are upper and lower bounds respectively, for the first-order phase transition line.
The sets of points are well fitted by a straight line of slope $1/2$.} \label{fig:log-log}
\end{centering}
\end{figure}

We show the value of $G_0$ as a function of $\beta$ on Figure \ref{fig:Fofbeta}.
The first-order phase transition ($\beta = \beta_c$) is found as $G_0(A \neq 0)$ vanishes.
We compute the line $\beta_c(a_4 > 0)$ using continuation on $\beta$ and on $a_4$.
Just like $16a = 3 b^2$ is the first-order phase transition for the normal form $s_{a,b}$
(Figure \ref{fig:tricritical}), we recover the scaling
\begin{equation}\nonumber
a_4 \sim \left(1 + \frac{\beta}{\lambda_1}\right)^{1/2}
\end{equation}
on the first-order phase transition line, as shown Figure \ref{fig:log-log}.

\begin{figure}[htp]
\begin{centering}
\input{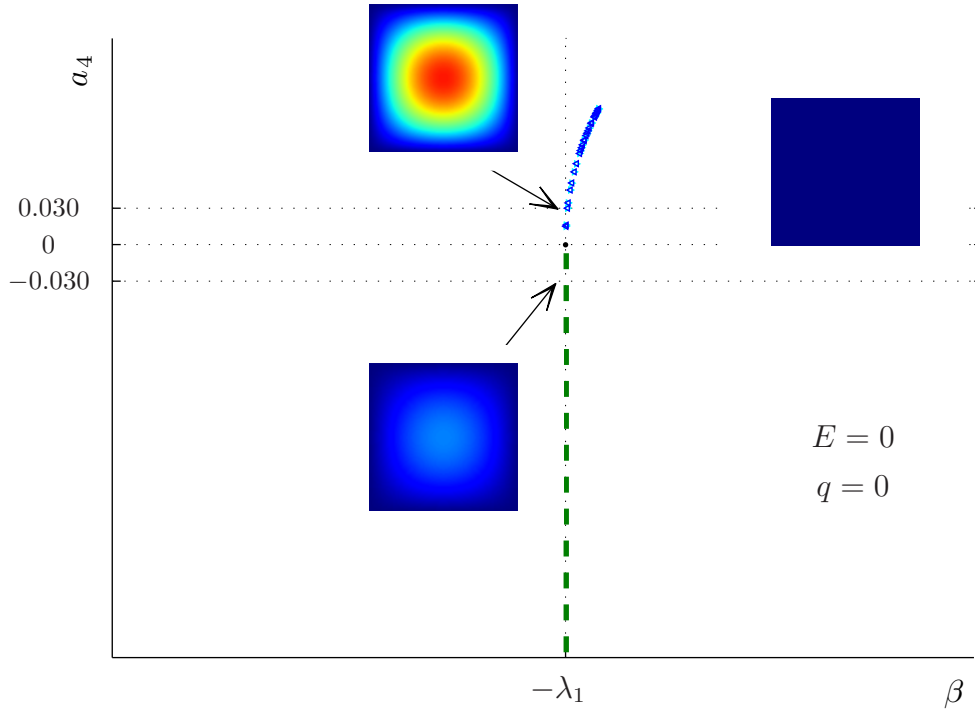}
\caption{Phase diagram in the vicinity of the (grand canonical, $\gamma = 0$) tricritical point
---black dot at $(-\lambda_1, 0)$, where the second-order phase transition ---green dashed line at $(-\lambda_1, a_4 < 0)$---
and the first-order phase transition ---line $(\beta_c(a_4), a_4 > 0)$ between the light blue dot series and
the dark blue dot series--- meet.  Insets show vorticity fields at $(\beta_c-(0.006\pm0.001), 0.030)$ and at $(-\lambda_1-(0.006\pm0.001), -0.030)$;
color scale ranges from $0$ to $0.6$
(from blue to red).} \label{fig:ph-diag}
\end{centering}
\end{figure}

\begin{figure}[htp]                                               
\begin{centering}
\includegraphics[width=4.0in]{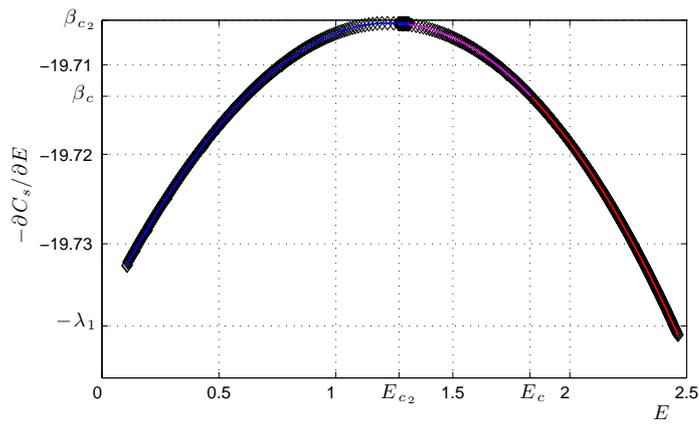}\\
\caption{\small Caloric curve for $a_4 = 0.030$.} \label{fig:caloric}
\end{centering}
\end{figure}

The phase diagram in the grand canonical ensemble ($\gamma=0$) is shown Figure \ref{fig:ph-diag}.
Figure \ref{fig:caloric}, we show the caloric curve for a positive value of $a_4$.

\subsection{Triple point for $\gamma \neq 0$\label{sub:asymm}}

We show the effect of having topography not orthogonal to the largest-scale mode $e_1$
and of having $\gamma \neq 0$: this is the general case for the grand canonical problem.
Then, the normal form reads
\begin{align}\nonumber
G(A) =& \left(\langle e_1 \rangle \gamma -\frac{\beta}{\lambda_1} h_1\right)A +
\frac{1}{2} \left(1 + \frac{\beta}{\lambda_1}\right) A^2 - \frac{a_4}{4} \langle e_1^4 \rangle A^4 +\notag\\
&+ O(A^3 \gamma, A^6, A^3 \gamma^3, A^4 \gamma^2).\notag
\end{align}
We see that the effect is that of breaking the $A \mapsto -A$ symmetry of $G(A)$, which is
a normal form for the grand canonical potential (to be minimized).
We may take $h_1 = 0$ without loss of generality, because the effect of $h_1 \neq 0$ is qualitatively
encompassed by $\gamma \neq 0$.
\par\smallskip\indent
Since the second-order phase transition we had originated from the $A \mapsto -A$ symmetry, we lose it
in the general case $\gamma \neq 0$.  Therefore, the tricritical point is lost.
We are left with a critical point, when the first-order phase transition survives.  It does so for small enough
$|\gamma|$.  It simply gets shifted in phase diagram $(\beta, a_4)$: now, $\beta_c$ depends on both
$a_4$ and $\gamma$.  We illustrate this, at given small $a_4 > 0$, in figure \ref{fig:ph-diag}.  
In the grand canonical ensemble, we have a triple point in phase diagram $(\beta, \gamma)$.

\begin{center}
\begin{figure}[htp]
\input{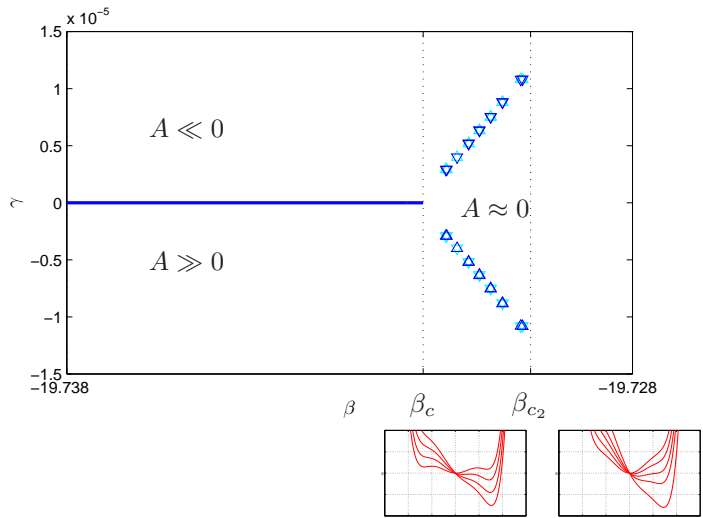}
\caption{Phase diagram in dual space $(\beta, \gamma)$ at $a_4 = 0.015$, displaying a first-order transition line
at $\gamma = 0$
up to $\beta_c > -\lambda_1$, splitting into two first-order transition lines ($\gamma \mapsto -\gamma$ symmetry) for
$\beta\!\in\: ]\beta_c, \beta_{c_2}[$ (insets below the phase diagram show sketches of a sixth-order normal form for $G(A)$,
when $a_4 > 0$: different curves on each diagram correspond to $\gamma = \{-0.001, -0.01, -0.02, -0.03, (-0.05)\}$
from top to bottom, when looked at in domain $A \geq 0\,;$ lhs is for $-\lambda_1 < \beta < \beta_c\,,$ rhs is for
$\beta_c < \beta < \beta_{c_2}$).  Values of $A$ for the solution states are shown
in the different regions of dual space.} \label{fig:bet-gam-nz} 
\end{figure}
\end{center}

\subsection{Solutions for $\Gamma \neq 0$ and $\tau = 2$}

\begin{figure}[htp]
\begin{centering}
\includegraphics[width=4.0in]{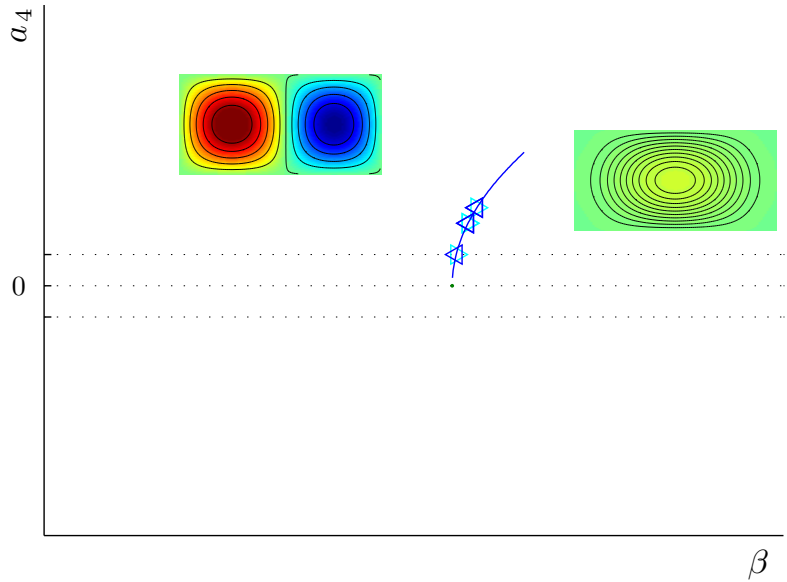}
\caption{Phase diagram in the canonical ensemble (circulation equal to $0.01$) for a rectangle of aspect ratio $2$
(case {\bf ii)}).  The blue curve plots $a_4 = 0.13 \sqrt{\lambda'_1+\beta}$; the $0.13$ prefactor was chosen so as to fit the three first-order phase transition points computed from numerical continuation.  This first-order phase transition line ends at a
second-order phase transition point ---green dot at $(-\lambda'_1, 0)$.  Insets show vorticity fields at $(\beta_c^{-}, 0.030)$ and at $(\beta_c^{+}, -0.030)$; color scale ranges from $-0.5$ to $0.5$ (from blue to red); the black contours are ten iso-vorticity lines on each plot.} \label{fig:ph-diag-canii}
\end{centering}
\end{figure}

Figure \ref{fig:ph-diag-canii}, we show the computed solution for low but nonzero circulation (canonical ensemble)
and $\tau = 2$ ---elongated rectangle, case \textbf{ii)}.

\newpage
\section{Physical applications\label{sec:applns}}

First, let us mention that there is a theoretical interest for a classification
of phase transitions.  Notably, two-dimensional flows are long-range interacting
systems (see subsection \ref{sub:ens-ineq}). The nice thing about these systems is that
theoretical results for one of them is relevant and useful to the others. So our
general results can extend to other systems with long-range interactions.
\par\medskip\indent
Let us come back to the geophysical motivation though.
The quasi-geostrophic equations serve as a simple model
for the motion of atmospheric or oceanic flows.
The inviscid quasi-geostrophic model is thus an appropriate model for geophysical and experimental flows
on time scales much less than the dissipation time scale, but large enough
for turbulent mixing to have operated as much as allowed by the constraint of
energy conservation.  
For instance, such an equilibrium approach led to a successful description
of the self-organization of Jupiter's atmosphere;
especially, it led to a realistic model for Jupiter's Great Red Spot and other vortices~\cite{Bouchet_Sommeria:2002_JFM}.
\par\smallskip\indent
Recently, it was shown that ocean currents, such as the Kuroshio (in the north Pacific ocean, off Japan)
or the Gulf Stream, may also be understood as equilibria of the inertial dynamics,
in very simple ocean models~\cite{Venaille_Bouchet_2011_JPO}.
Naturally, this conservative theory ignores all effects due to forcing and dissipation, which are
present in any real flow.  Still, a recent work showed that the inertial description
of equilibria is fundamental and relevant even in the presence of forcing and dissipation
\cite{Bouchet_Simonnet_2009PRL}.
Fluctuations are responsible for the phenomenon of transitions between two equilibria
(bistability).
\par\smallskip\indent
The barotropic quasi-geostrophic model is also relevant to the description
of experimental flows, such as the approximation of fluid dynamics when three-dimensional motion
is constrained by a strong transverse field (e.g., rotation)
or takes place in geometries of small (vertical-to-horizontal) aspect ratio.
Experiments with fluid in a rotating annulus, using a forcing mechanism, enable to
produce a zonal (azimuthal) jet subject to the Coriolis force.
In such fast-rotating tanks equipped with ridges at the bottom (mimicking topography),
flow patterns identified as `zonal' and `blocked' states are observed.  In addition,
transitions between the two states are found in a certain range of forcings
(tank rotation and pumping rate)
\cite{Tian_Weeks_etc_Ghil_Swinney_2001_JFM_JetTopography}.
In the blocked state, streamlines tend to follow topography contours.
\par\smallskip\indent
This bistability is reminiscent of the phenomenon of atmospheric `blocking':
On interannual time scales, large anticyclones form in the Northern Hemisphere,
blocking and deflecting the nearly zonal flow (following
latitude circles) \cite{Weeks_Tian_Urbach_Ide_Swinney_Ghil_1997}.  Analogous configurations
are observed in the north Pacific ocean, where the Kuroshio Extension forms an eastward
mid-basin jet.  The Kuroshio is seen to oscillate
between an intense jet-like (zonal) state and a weaker meandering (blocked) state.  
\par\medskip\indent
Our intuition is that the qualitatively different states predicted by our
phase diagrams could be related to the different regimes observed in geophysical
flows, providing that realistic geometries are considered (annular domain for example,
coastline geometry, bottom topography).
Fluctuations would be responsible for the transitions between different equilibria.

\newpage
\section{Appendix: Grand canonical solutions\label{app:gdcan-sol}}

Let $M_g = \{q' | \langle q' e_1 \rangle = 0\}$. We may decompose $h = h_1 e_1 + h'$,
with $h' \in M_g$.  Through $\Delta \psi = q - h$, we have $\psi = -\frac{(A-h_1)}{\lambda_1} e_1 + \psi'$,
with $\psi' \in M_g$.
Let us denote the Gibbs free energy \eqref{G_functional} with $\gamma = 0$ by ${\mathcal G}_0$.
\begin{equation}\nonumber
\mathcal{G}_0[q] = \mathcal{G}_0[A,q'] = \int_{\mathcal D} \left[s(A e_1 + q') - \frac{\beta}{2}\psi' (q'-h') \right]
+ \frac{\beta}{2} \frac{(A-h_1)^2}{\lambda_1}.
\end{equation}
The second-order variation of ${\mathcal G}_0$ with respect to $q'$ is
\begin{equation}\label{secondvarF}
\delta^2 \mathcal{G}_0[A,q'] = \int_{\mathcal D} s''(A e_1 + q') \delta q'^2 - \beta \int_{\mathcal D} \delta\psi' \delta q'.
\end{equation}
It is straightforward to prove a generalization of the Poincar\'e inequality in the subspace $M_g$
(any $q' \in M_g$ may be written $q' = \sum_{i\geq2} q_i e_i$),
which yields, for $\beta < 0$, the inequality
\begin{equation}
\delta^2 \mathcal{G}_0[A,q'] \geq \left(s^g_A+ \frac{\beta}{\lambda_2} \right) \int_{\mathcal D} \delta q'^2, \notag
\end{equation}
where $s^g_A := \min_{{\bf r} \in {\mathcal D}} \{\min_{q'} s''(A e_1+q'({\bf r}))\}$.
Therefore, if $\beta > - s^g_A \lambda_2$, $\mathcal{G}_0$ is convex with respect to $q'$ and we denote by $q'_{eq}$
the unique solution to the minimization problem
\begin{align}\label{G0A}
G_0(A) =& \min_{q'} \mathcal{G}_0[A,q'] = \int_{\mathcal D} \left[s(A e_1 + q'_{eq}) - \frac{\beta}{2}\psi'_{eq} (q'_{eq}-
h') \right] +\notag\\
&+ \frac{\beta}{2} \frac{(A-h_1)^2}{\lambda_1}.
\end{align}
\par\smallskip\indent
For $\beta > - s^g_A \lambda_2$, $q'_{eq}$ is the unique critical point of $\mathcal{G}_0$ with respect to $q'$.  It satisfies
\begin{equation}\nonumber
\int \big(s'(A e_1 + q'_{eq}) - \beta\psi'_{eq} \big) \delta q' = 0 \quad \text{for all} \; \delta q' \in M_g,
\end{equation}
therefore there exists $\alpha_g \in \mathbb{R}$ such that
\begin{equation}\label{hcrit}
s'(A e_1 + q'_{eq}) - \beta\psi'_{eq} = \alpha_g e_1.
\end{equation}
We compute the solution to \eqref{hcrit} perturbatively around $(A, q') = (0,0)$, in order to obtain an asymptotic
expansion for $G_0(A)$ around $A = 0$, and hence determine
the type of phase transitions to expect in the vicinity of $(\beta \leq - s''_m \lambda_1\,, a_4 = 0)$.
Remark that if $a_{2 n} \leq 0$ for all $n \geq 2$, then $s''_m = \min_q \{1 - 3 a_4 q^2 - o(q^3)\} = 1$
and $s''_m \to 1$ as $(a_4, q) \to (0,0)$; $s(q)$ is said to be strongly
convex\footnote{\url{http://en.wikipedia.org/w/index.php?title=Convex_function&oldid=466661785\#Strongly_convex_functions}
Accessed January 5, 2012.}.
\par\medskip\indent
We have the Taylor expansion $s'(q) = q - a_4 q^3 - a_6 q^5 + o(q^6)$.  Substituting this expression into \eqref{hcrit}, and projecting \eqref{hcrit} orthogonally onto $M_g$
(projection of $x$ being denoted by $P(x) := x - \langle x e_1 \rangle e_1$), we get
\begin{align}
&q'_{eq} - a_4 A^3 P(e_1^3) - 3 a_4 A^2 P(e_1^2 q'_{eq}) - 3 a_4 A P(e_1 q'^2_{eq}) - a_4 P(q'^3_{eq}) + \notag \\
& + O(A^5, A^4 q'_{eq}, A^3 q'^2_{eq}, A^2 q'^3_{eq}, A q'^4_{eq}, q'^5_{eq}) = \beta\psi'_{eq}. \notag 
\end{align}
At lowest order in the asymptotic expansion of $q'_{eq}$ ($q'_{eq} = q'_0 +$ higher powers of $A$),
we have
\begin{equation}\nonumber
q'_0 - \beta \psi'_0 = \Delta \psi'_0 - \beta \psi'_0 = a_4 A^3 P(e_1^3);
\end{equation}
the linear operator $\mathcal{L}_\beta : \psi' \mapsto \Delta \psi' -\beta \psi'$ is invertible in the subspace $M_g$
for $\beta$ in the vicinity of $-\lambda_1$.
Thus, we get
\begin{equation}
\left\{\begin{aligned}
\psi'_0 &= a_4 A^3 \mathcal{L}_\beta^{-1} P(e_1^3) =: \tilde{\psi}'_0 A^3, \\
q'_0 &= a_4 A^3 \Delta \mathcal{L}_\beta^{-1} P(e_1^3) =: \tilde{q}'_0 A^3.
\end{aligned}
\right.
\end{equation}
Now, we compute the asymptotic expansion of $G_0(A)$ using this perturbative result: substituting
$q'_{eq} = \tilde{q}'_0 A^3 + o(A^3)$ into \eqref{G0A}, we get
\begin{align}\label{GofA-g}
G_0(A) &= - \frac{\beta}{\lambda_1} h_1 A + \frac{1}{2} \left(1 + \frac{\beta}{\lambda_1}\right) A^2 - \frac{a_4}{4} \langle e_1^4 \rangle A^4 + \\
&- \left[a_4 \int \tilde{q}'_0 e_1^3+ \frac{1}{2} \int \tilde{q}'^2_0 - \frac{\beta}{2} \int \tilde{\psi}'_0 \tilde{q}'_0 + \frac{a_6}{6} \langle e_1^6 \rangle \right] A^6 
+ o(A^6). \notag
\end{align}
The parity of $G_0(A)$ is broken by $h_1 \neq 0$.  Let us take $h_1 = 0$ until further notice.  Note that
up to quartic order, only mode $e_1$ contributes ---the perturbation $q'_{eq}$ contributes only from order $6$ and up.

\section{Appendix: Canonical solutions\label{app:canon-sol}}

Recall
\begin{equation}\nonumber
F(\beta, \Gamma) = \min_{q} \left\{\mathcal{F}[q] = \int_{\mathcal D} s(q) + \beta \mathcal{E}[q] \ |
\ {\mathit \Gamma}[q] = \Gamma \right\}.
\end{equation}
We reduce the set of independent variables to $\{ q_i \}_{i \geq 2}$, as in
\cite{Venaille_Bouchet_PRL_2009}.  Since
\begin{equation}\label{eq:circul-constr}
\Gamma = \sum_i q_i \langle e_i \rangle\,, 
\end{equation}
we may decompose
\begin{equation}\label{q-tilde}
q = \frac{\Gamma}{\langle e_1 \rangle} e_1 + \sum_{i \geq 2} q_i \left(e_i - \frac{\langle e_i \rangle}{\langle e_1 \rangle}
e_1\right) =: \frac{\Gamma}{\langle e_1 \rangle} e_1 + q_c\,,
\end{equation}
so as to consider the minimization of $\mathcal F$ with respect to $q_c$:
\begin{equation}\label{var-pb-canon}
\min_{q} \left\{\mathcal{F}[q] \ | \ {\mathit \Gamma}[q] = \Gamma \right\} = \min_{q_c}
\left\{\mathcal{F} \left[\frac{\Gamma}{\langle e_1 \rangle} e_1 + q_c\right] \right\}.
\end{equation}
Let $M_c 
= \{ q_c,\, \text{determined by} \ \{ q_i \}_{i \geq 2} | \langle q_c \rangle = 0 \}$.
$M_c$ is a subspace complementary to the line spanned by $e_1$.  $M_c$ is a subspace
orthogonal to $1$.  Canonical solutions live in $M_c$.
\par\smallskip\indent
Note
\begin{equation}\nonumber
s_\Gamma := \min_{{\bf r} \in {\mathcal D}} \left\{ \min_{q_c \in M_c}
s''\left(\frac{\Gamma}{\langle e_1 \rangle} e_1 + q_c\right)\right\},
\end{equation}
and consider $\delta^2 \mathcal{F}$,
the second-order variation of $\mathcal{F}$ with respect to $q_c$. 
In appendix \ref{Poinc-cani}, we prove a generalization of the Poincar\'e inequality in the subspace $M_c$, leading
to $\delta^2 \mathcal{F} \geq (s_\Gamma + \beta/\lambda_{c}) \langle \delta q_c^2\rangle$.
We have introduced $\lambda_c := \min \{\lambda_*, \lambda'_1\}$, corresponding
to the vanishing of the quadratic part of $\mathcal{F} [\Gamma e_1/\langle e_1 \rangle + q_c]$
(denoted by $\mathcal{Q}_{\mathcal{F}}$)
at $\beta = -\lambda_{c}$.  The space $M_c$ is a direct sum of the subspace
generated by eigenmodes of zero domain average, $\{e'_i\}_{i \geq 1}$, and the subspace
generated by all the other modes.  In the former subspace, $\mathcal{Q}_{\mathcal{F}}$
vanishes at $\beta = -\lambda'_1$ along $e'_1$.  In the latter subspace,
$\mathcal{Q}_{\mathcal{F}}$ vanishes at $\beta = -\lambda_*$ along $e_*$, where $\lambda_*$
is the smallest value of $-\beta$ such that
\begin{equation}\label{eq:chap2-sum}
\hat{f}(\beta) = - \sum_{i \geq 1} \frac{\lambda_{i} \langle e_{i} \rangle^2}{\lambda_{i} +\beta} = 0.
\end{equation}
The interested reader can find details about the above function in
\cite{Chavanis_Sommeria_1996JFM}.
Anyhow, there are no phase transitions in the canonical ensemble for $\beta > - s_\Gamma \lambda_{c}$.
\par\medskip\indent
In appendix \ref{app:LSred}, we detail the computation of the $\Gamma = 0$ solutions.
For circulation $\Gamma$, the expression of the `linear' solution ($a_4 = 0$) is
\begin{align}
q(\beta > -\lambda_c, \Gamma) &= -\frac{\Gamma}{\hat{f}(\beta)} \sum_{i \geq 1} \frac{\lambda''_i \langle e''_i \rangle}{\lambda''_i + \beta} e''_i , \notag\\
q(\beta = -\lambda_c, \Gamma) &= -\frac{\Gamma}{\hat{f}(\beta)} \sum_{i \geq 1} \frac{\lambda''_i \langle e''_i \rangle}{\lambda''_i + \beta} e''_i \pm A e_c.
\label{q_0lin}
\end{align}
We can see that a nonzero circulation will introduce a symmetry breaking into
the normal form \eqref{FofAi}--\eqref{FofAii}.
We consider a small circulation $|\Gamma|$, for the description to remain close to the zero-circulation case.  Also, this is
required by the low-energy limit and the vicinity of $\beta = \beta_c$.
Because of the $A \mapsto -A$ symmetry breaking, due to $\Gamma \neq 0$, the second-order phase transition vanishes, leaving a phase diagram with a critical point. 

\section{Appendix: Poincar\'e inequality in the canonical ensemble\label{Poinc-cani}}

In this appendix, we prove a generalization of the Poincar\'e inequality to the case with fixed circulation, i.e.,
in the subspace $M_c$.
\par\smallskip\indent
Let $\tilde{q} \in M_c$.  Then, $\delta \tilde{q} = \sum_{i \geq 2} \delta q_i \big(e_i - \frac{\langle e_i \rangle}{\langle e_1 \rangle} e_1\big)$\\
and
$\delta \tilde{\psi} = -\sum_{i \geq 2} \delta q_i \big(\frac{e_i}{\lambda_i} - \frac{\langle e_i \rangle}{\langle e_1 \rangle} \frac{e_1}{\lambda_1}\big)$.  We have
\begin{align}
\int_{\mathcal D} \delta \tilde{q}^2 &= \sum_{i \geq 2} \delta q^2_i
+ \frac{1}{\langle e_1 \rangle^2} \sum_{i,j \geq 2} \langle e_i \rangle \langle e_j \rangle \delta q_i \delta q_j, \notag \\
- \beta \int_{\mathcal D} \delta\tilde{\psi} \delta \tilde{q} &= \beta \sum_{i \geq 2} \frac{\delta q^2_i}{\lambda_i}
+ \frac{\beta}{\lambda_1 \langle e_1 \rangle^2} \sum_{i,j \geq 2} \langle e_i \rangle \langle e_j \rangle \delta q_i \delta q_j. \notag
\end{align}
Now,
\begin{align}
\int_{\mathcal D} \delta \tilde{q}^2 - \beta \int_{\mathcal D} \delta\tilde{\psi} \delta \tilde{q} &= \sum_{i \geq 1} \left(1 + \frac{\beta}{\lambda'_i} \right) \delta q'^{2}_i + \notag \\
 &+ \sum_{i,j \geq 2} \left[\delta_{i j} \left(1 + \frac{\beta}{\lambda''_i}\right) + \left(1 + \frac{\beta}{\lambda''_1}\right)\frac{\langle e''_i \rangle
\langle e''_j \rangle}{\langle e_1 \rangle^2}\right] \delta q''_i \delta q''_j \notag
\end{align}
is positive definite if and only if $\beta > -\min \{\lambda'_1, \lambda^* \} = -\lambda_c$.
\par
Since $- \beta \int_{\mathcal D} \delta\tilde{\psi} \delta \tilde{q} \geq -\beta/\beta \int_{\mathcal D} \delta \tilde{q}^2$ for all $\beta \in [-\lambda_c\,, 0[\,,$
then the best (greatest) lower bound that we can obtain is
\begin{equation}\nonumber
- \beta \int_{\mathcal D} \delta\tilde{\psi} \delta \tilde{q} \geq \frac{\beta}{\lambda_c} \int_{\mathcal D} \delta \tilde{q}^2.
\end{equation}

\section{Appendix: Lyapunov--Schmidt reduction in the canonical ensemble\label{app:LSred}}
In this appendix, we derive the phase diagram for the canonical solutions at zero circulation.
Consider the following canonical variational problem (let us drop the $\mathcal D$ subscript in the integral notation):
\begin{equation}\nonumber
\min_q = \left\{ \int s(q) - \frac{\beta}{2} \int q \psi \ | \int q = 0 \right\}.
\end{equation}
A critical point is $q$ such that
\begin{equation}\nonumber
\int s'(q) \delta q - \beta \int \psi \delta q = 0 \ \text{for all} \ \delta q \in {\mathcal Q} \ \text{such that} \ \int \delta q = 0,
\end{equation}
or, equivalently, using the Lagrange multiplier rule,
\begin{equation}\label{eq:canon}
\tilde{f}(q, \gamma; \beta) := \left\{\begin{aligned}
\tilde{f}_1(q, \gamma; \beta) &= s'(q) - \beta \psi + \gamma = 0,\\
\tilde{f}_2(q, \gamma; \beta) &= \int q = 0,
\end{aligned}
\right.
\end{equation}
where $\gamma \in \mathbb{R}$ is the Lagrange parameter associated with the conservation of
(zero) circulation.
\par\indent
The system \eqref{eq:canon} is to be solved in the variables $(q, \gamma)$, while the bifurcation parameter is $\beta \in \mathbb{R}$.  Let us denote the variable by $X = (q, \gamma)$ and the
variable space by $E$.  Please do not get this notation mixed up with the energy, which we never mention in this
appendix.  $\tilde{f}$ maps $E \times \mathbb{R}$ into $E$.
For any $\beta \in \mathbb{R}$, we have the trivial solution $X = 0$.
We want to determine the bifurcations, which the system may undergo, from this trivial solution.
\par\medskip\indent
For a bifurcation to occur, the Jacobian matrix of \eqref{eq:canon} has to become singular, i.e.,
there must exist a nontrivial vector $u_c = (q_c, \gamma_c) \in E$ such that $D_X \tilde{f}(0; \beta)[u_c] = 0$
for a certain $\beta = \beta_c$.
We have
\begin{equation}\label{canon-2}
D_X \tilde{f}(0; \beta)[u_c] = \left(\begin{aligned}
&s''(0) \delta q_c - \beta \delta \psi_c + \delta \gamma_c \\
&\int \delta q_c
\end{aligned}
\right) \in E,
\end{equation}
with $\Delta \psi_c = q_c$.
Let us endow $E$ with the scalar product
$(\cdot | \cdot)$, defined as follows: for $X_k = (q_k, \gamma_k) \in E$, $k =\{ 1, 2\}$,
\begin{equation}
(X_1|X_2) = \langle q_1 q_2 \rangle + \gamma_1 \gamma_2 = \int (q_1 q_2) + \gamma_1 \gamma_2.
\end{equation}
A complete orthonormal basis for $E$ is $\{u_{i}\}_{i \geq 0}$, where $u_0 = (0,1)$ and $u_i = (e_i, 0)$
for $i \geq 1$.  $D_X \tilde{f}(0; \beta)$ is self-adjoint since
\begin{align}\nonumber
(X_1|D_X \tilde{f}(0; \beta)[X_2]) &= \int q_1 (s''(0) \delta q_2 - \beta \delta \psi_2 + \delta \gamma_2) 
+ \gamma_1 \int \delta q_2 \notag \\
&= s''(0) \langle q_1 \delta q_2 \rangle - \beta \langle q_1 \delta \psi_2 \rangle + \langle q_1 \delta \gamma_2 \rangle
+ \langle \gamma_1 \delta q_2 \rangle\notag \\
&= (D_X \tilde{f}(0; \beta)[X_1]|X_2), \notag
\end{align}
so $D_X \tilde{f}(0; \beta)$ may be diagonalized in $\{u_{i}\}_{i \geq 0}$, and its eigenvalues are real.
The equalities $\langle q_1 \delta q_2 \rangle = \langle q_2 \delta q_1 \rangle$, $\langle q_1 \delta \psi_2 \rangle =
\langle q_2 \delta \psi_1 \rangle$, and so on, come from the Euclidean-ness of $E$.
Indeed, let $k = \{ 1, 2 \}$ and $q_k = \sum_i q_{k,i} e_i$.  The tangent vector $\delta q_k = \sum_i \delta q_{k,i} e_i$
is along $q_k$, so for all $i \geq 1$,
$\delta q_{k,i} = a_k q_{k,i}$.  Now, $a_1 = a_2$ because $q_1$ and $q_2$, belonging to the same space,
must be mapped onto their tangent space with the same coefficient.
Decomposing the variables in the Laplacian eigenbasis $\{ e_i \}_{i \geq 1}$,
\begin{equation}
\delta q = \sum_{i \geq 1}\delta q_i e_i \ ; \quad \delta \psi = -\sum_{i \geq 1}\frac{\delta q_i}{\lambda_i} e_i \ ; \quad
\delta \gamma = \delta \gamma \sum_{i \geq 1}\langle e_i \rangle e_i,
\end{equation}
it is readily seen that $u_c$ is either along $(e'_i, 0)$ at $\beta = -s''(0) \lambda'_i$,
or along $(e_*, 1)$ at $\beta = -s''(0) \lambda_*$.  Indeed, we identify
\begin{equation}\nonumber
\delta q_i = -\delta \gamma \frac{\lambda_i \langle e_i \rangle}{s''(0) \lambda_i + \beta} \quad
\text{for all} \ i \geq 1,
\end{equation}
and we require
\begin{equation}\nonumber
\langle \delta q \rangle = \sum_{i \geq 1} -\delta \gamma \frac{\lambda_i \langle e_i \rangle^2}{s''(0) \lambda_i + \beta}
= \frac{\delta \gamma}{s''(0)} \hat{f}\left(\frac{\beta}{s''(0)}\right) = 0,
\end{equation}
where the $\hat{f}$ function is \eqref{eq:chap2-sum}.  We have noted
\begin{equation}\nonumber
e_* := -\sum_{i \geq 1} \frac{\lambda_i \langle e_i \rangle}{s''(0)\lambda_i - \lambda^*} e_i .
\end{equation}
Note again that $\delta X$ belongs to the tangent space of $E$, but $E$ is Euclidean, so
$\{ X \in E \ | \ \delta X = a u_c, a \in \mathbb{R} \} = \{ X \in E \ | \ X = a u_c, a \in \mathbb{R} \}$.
\par\smallskip\indent
The first bifurcation, and hence, phase transition, to occur is found at $\beta_c = - s''(0) \lambda_c$
(considering a decreasing $\beta$).  Let
$u_c = \mathcal{N} (e_*, 1)$ in case \textbf{i)}, $u_c = (e'_1, 0)$ in case \textbf{ii)}.
$\mathcal{N}$ is the normalization factor $(\langle (e_*)^2 \rangle + 1)^{-1/2}$.
\par\medskip\indent
Let us denote the operator $D_X \tilde{f}(0; \beta_c)$ by $J$.  $J$ maps $E$ into $E$.
Let $E_c$ be the null space (kernel) of $J$.  It is the subspace generated by $u_c$; it is $1$-dimensional in
$E$ (it is a line). 
Let us show that the range of $J$ is orthogonal to $E_c$.  This is the case if and only if
$(Y|u_c) = 0$ for any $Y$ in the range of $J$.
\begin{itemize}
\item[\textbf{i)}]Let us show that
\begin{equation}\nonumber
\langle (s''(0) \delta q - \beta_c \delta \psi + \delta \gamma) e_* \rangle + \langle \delta q \rangle = 0.
\end{equation}
Let $\psi_*$ be the vector such that $\Delta \psi_* = e_*$ and $\psi_* = 0$ on $\partial \mathcal{D}$.
We have $\langle \psi e_* \rangle =
\langle q \psi_* \rangle $ (straightforward when decomposing in the Laplacian eigenbasis), so
\begin{align}
\langle (s''(0) \delta q - \beta_c \delta \psi + \delta \gamma) e_* \rangle + \langle \delta q \rangle
= \langle (s''(0) e_* - \beta_c \delta \psi_* + 1) \delta q \rangle = 0. \label{eq:parcasei}
\end{align}
Indeed, the last parenthesed term is the first component of $J u_c$ ($J u_c = 0$).
\item[\textbf{ii)}]Let us show that
\begin{equation}\nonumber
\langle (s''(0) \delta q - \beta_c \delta \psi + \delta \gamma) e'_1 \rangle = 0.
\end{equation}
We have $\beta_c = - s''(0) \lambda'_1$, so
\begin{equation}\nonumber
\langle (s''(0) \delta q - \beta_c \delta \psi + \delta \gamma) e'_1 \rangle =
\left\langle \left(s''(0) e'_1 + \frac{\beta_c}{\lambda'_1}e'_1 \right) \delta q \right\rangle = 0.
\end{equation}
\end{itemize}
Therefore, the kernel of $J$ is orthogonal to the range of $J$.
We can then apply classical bifurcation theorems~\cite{chow1982methods}.
Let $E_1$ be the orthogonal complementary subspace to $E_c$ in $E$ ($E_1$ is the range of $J$).
There exist $\tilde{X}(A, \beta) \in E_1$ ($A \in \mathbb{R}$) such that $\tilde{X}(0, \beta_c) = 0$ and
$\frac{\partial \tilde{X}}{\partial A}(0, \beta_c) = 0$, so that we may decompose the variable $X$ as follows:
\begin{equation}
X = X(A, \beta) = A u_c + \tilde{X}(A, \beta) = A u_c + \big(\tilde{q}(A, \beta), \tilde{\gamma}(A, \beta)\big).
\end{equation}
We will also use the notation $\tilde{\psi}$ for the vector in $\mathcal Q$ such that $\Delta \tilde{\psi} = \tilde{q}$.
Besides, there exists a projector $Q : E \to E_1$, $Q X = X - (X|u_c)u_c$ such that
$Q \tilde{f} (X; \beta) = Q \tilde{f}(A u_c + \tilde{X}; \beta) = 0$ for all $A, \beta \in \mathbb{R}$.
Now,
\begin{equation}\nonumber
\tilde{f}(A u_c + \tilde{X}; \beta) = Q \tilde{f}(A u_c + \tilde{X}; \beta) + (\tilde{f}(A u_c + \tilde{X}; \beta)|u_c)u_c
\end{equation}
so the bifurcation problem \eqref{eq:canon} is equivalent to (reduces to) the scalar problem
\begin{equation}\nonumber
h(A, \beta) := (\tilde{f}(A u_c + \tilde{X}(A, \beta); \beta)|u_c) = 0
\end{equation}
(Lyapunov--Schmidt reduction).  From the normalization of $u_c$, we have $f(A u_c + \tilde{X}; \beta) = h(A, \beta) u_c$.
Explicitly, this writes
\begin{itemize}
\item[\textbf{i)}]
\begin{equation}\label{alongi}
\left\{\begin{aligned}
&s'(A \mathcal{N} e_* + \tilde{q}) - \beta (A \mathcal{N} \psi_* + \tilde{\psi}) + A\mathcal{N} + \tilde{\gamma} =
\mathcal{N} h(A, \beta) e_* , \\
&\int A \mathcal{N} e_* + \tilde{q} = \mathcal{N} h(A, \beta) ;
\end{aligned}
\right.
\end{equation}

\item[\textbf{ii)}]
\begin{equation}\label{alongii}
\left\{\begin{aligned}
&s'(A e'_1 + \tilde{q}) + \beta (\frac{A}{\lambda'_1} e'_1 - \tilde{\psi}) + \tilde{\gamma} =
h(A, \beta)e'_1 , \\
&\int A e'_1 + \tilde{q} = 0 .
\end{aligned}
\right.
\end{equation}
\end{itemize}
We may notice that for $(A, \tilde{X}, h)$ solution, $(-A, -\tilde{X}, -h)$ is also a solution, so that $h$ and $\tilde{X}$ are
odd in $A$.  Therefore $\frac{\partial^2 h}{\partial A^2}$ and $\frac{\partial^2 \tilde{X}}{\partial A^2}$ are also odd in $A$,
and so on. 
\par\smallskip\indent
We know that $F$ is even in $A$.  We have $F(A=0) = 0$, so the lowest order of $F$ is quadratic.  We determine the successive coefficients (of each power of $A$) in $F$ from its successive derivatives w.r.t. $A$, evaluated at $A=0$.
Because $\langle \tilde{q} e_c \rangle = 0$, we also have $\langle \frac{\partial\tilde{q}}{\partial A} e_c \rangle = 0$, and
so on with all the derivatives with respect to the scalar $A$.  All these properties lead to drastic simplifications in
the computation of
$\frac{\mathrm{d}^2 F}{\mathrm{d} A^2}(A=0)$ and $\frac{\mathrm{d}^4 F}{\mathrm{d} A^4}(A=0)$, leaving us with
\begin{itemize}
\item[{\bf i)}]
\begin{equation}\label{FofAi}
F(A) = \frac{1}{2} \frac{\langle e_*^2 \rangle}{\langle e_*^2 \rangle+1}\left(s''(0) + \frac{\beta}{\lambda_*}\right) A^2
- \frac{\langle e_*^4 \rangle}{(\langle e_*^2 \rangle+1)^2}\frac{a_4}{4}A^4 + o(A^5);
\end{equation}
\item[{\bf ii)}]
\begin{equation}\label{FofAii}
F(A) = \frac{1}{2} \left(s''(0) + \frac{\beta}{\lambda'_1}\right) A^2 - \frac{a_4}{4} \langle e^{'4}_1 \rangle A^4 + o(A^5).
\end{equation}
\end{itemize}
\par\medskip\indent
Bifurcation-wise, it is shown that
\begin{equation}\nonumber
h(0, \beta_c) = 0, \quad \frac{\partial h}{\partial A}(0, \beta_c) = 0, \quad \frac{\partial^2 h}{\partial A^2}(0, \beta_c) = 0,
\end{equation}
but 
\begin{equation}\nonumber
\frac{\partial^3 h}{\partial A^3}(0, \beta_c) \neq 0; \quad \sgn\left(\frac{\partial^3 h}{\partial A^3}(0, \beta_c) \right) = -\sgn(a_4).
\end{equation}
Therefore, the bifurcation will be determined (qualitatively) by the cubic nonlinearity of $h$ in $A$ (corresponding to the
quartic nonlinearity of $F$ in $A$, in the present paper).  The sign of $a_4$, i.e., the parameter for the nonlinearity in the $q-\psi$ relationship, determines the type of bifurcation at play:
\begin{itemize}
\item[--] If $a_4 < 0$, the pitchfork bifurcation is supercritical, giving a second-order phase transition.
\item[--] If $a_4 > 0$, the pitchfork bifurcation is subcritical, giving a first-order phase transition (the higher-order nonlinearities
yielding nontrivial branches beyond $\beta = \beta_c$, at $\beta < \beta_c$).
\end{itemize}

\bibliographystyle{plain}
\bibliography{short_ED2,Meca_Stat_Euler,Maths}

\end{document}